\documentclass[aps,pra,superscriptaddress,showpacs,twocolumn]{revtex4}

\usepackage{braket}
\usepackage{graphicx}
\usepackage{epstopdf}

\usepackage{subfigure}  

\usepackage[latin1]{inputenc}
\usepackage{amsmath,amsbsy,amsfonts,amssymb}
\usepackage{epsfig}
\usepackage{hyperref}
\newcommand{\wn}{\mbox{cm$^{-1}$}}
\newcommand{\Xpot}{$\mathrm{X}\,^1\Sigma^+_\mathrm{g}$~}
\newcommand{\exasym}{${^3P_1}$+${^1S_0}$~}

\begin{document}

\title{Ground-state properties of Ca$_2$ from narrow line two-color photoassociation}

\author{Evgenij Pachomow}
\affiliation{Physikalisch-Technische Bundesanstalt (PTB), Bundesallee 100, 38116 Braunschweig, Germany}
\author{Veit Peter Dahlke}
\affiliation{Physikalisch-Technische Bundesanstalt (PTB), Bundesallee 100, 38116 Braunschweig, Germany}
\author{Eberhard Tiemann}
\affiliation{Institut f\"ur Quantenoptik, Leibniz Universit\"at Hannover, Welfengarten 1, 30167 Hannover, Germany}
\author{Fritz Riehle}
\affiliation{Physikalisch-Technische Bundesanstalt (PTB), Bundesallee 100, 38116 Braunschweig, Germany}
\author{Uwe Sterr}
\affiliation{Physikalisch-Technische Bundesanstalt (PTB), Bundesallee 100, 38116 Braunschweig, Germany}

\date{\today}

\begin{abstract}
By two-color photoassociation of $^{40}$Ca four weakly bound vibrational levels in the Ca$_2$ \Xpot ground state potential were measured, 
using highly spin-forbidden transitions to intermediate states of the coupled system $^3\Pi_{u}$ and $^3\Sigma^+ _{u}$ near the ${^3P_1}$+${^1S_0}$ asymptote.
From the observed binding energies, including the least bound state, the long range dispersion coefficients $\mathrm{C}_6, \mathrm{C}_8,\mathrm{C}_{10}$ and a precise value for the s-wave scattering length of 308.5(50)~$a_0$ were derived. From mass scaling we also calculated the corresponding scattering length for other natural isotopes.
From the Autler-Townes splitting of the spectra, the molecular Rabi frequency has been determined as function of the laser intensity for one bound-bound transition. 
The observed value for the Rabi-frequency is in good agreement with calculated transition moments based on the derived potentials, assuming a dipole moment being independent of internuclear separation for the atomic pair model. \\
  
\end{abstract}


\maketitle

\section{Introduction}

Photoassociation (PA) of ultracold atoms is a valuable tool to accurately determine interatomic interactions, to create ultracold molecules \cite{jon06}, or to modify their scattering properties by optical Feshbach resonances \cite{chi10}. 
In the PA process two atoms collide in the presence of a light field, inducing transitions from the atom-atom scattering continuum to bound molecular states at relatively large internuclear distances $R$. 
Therefore, the PA process is sensitive to the long range parts of the interaction potentials. 
Two-color PA via an excited molecular state to states in the ground molecular potential allows probing  the interaction of two atoms in their electronic ground states. 
It has been used with great success for the investigation of binding energies and scattering lengths of alkaline atoms Li \cite{abr97}, Na \cite{abe99}, K \cite{wan00}, Rb \cite{tsa97}, Cs \cite{van04a}, 
and more recently for systems with two valence electrons  He$^*$ \cite{kim04a}, Sr \cite{esc08} and Yb \cite{kit08}. 

\begin{figure}[t]
\includegraphics[width=1\columnwidth]{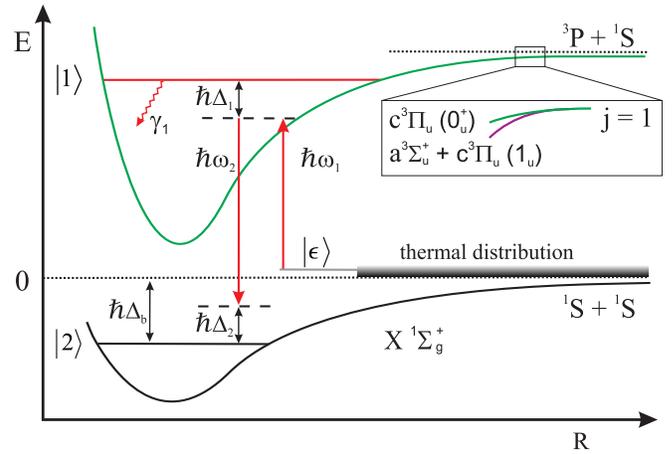}
\caption{Schematic overview of the two-color PA process featuring the involved molecular potentials of Ca$_2$, using intermediate states near the ${^1S_0}$ + ${^3P_1}$ asymptote ($\lambda=657$~nm).}
\label{fig:schema}
\end{figure}

In this paper we report on the first two-color PA of the alkaline-earth isotope $^{40}\mathrm{Ca}$. 
It has an electronic structure similar to Sr and Yb but for the atomic intercombination line a much narrower width of 374~Hz \cite{deg05a} compared to 7~kHz in Sr or 182~kHz in Yb. 
Due to its relatively simple electronic structure calcium is a good model system to investigate the theoretical predictions for lineshape and rate of collisions in the presence of laser fields given by the model of Bohn and Julienne \cite{boh96,boh99}.
We have measured the four least bound states ($ v = 40, J=0;~v = 39, J=0,2;~v = 38, J=2 $) in the $^{40}\mathrm{Ca}_2$ \Xpot ground state potential using strongly spin-forbidden intercombination transitions to intermediate states near the $^1S_0$ + $^3P_1$ asymptote (Fig.\ \ref{fig:schema}).

Our measurements complement the short range ground-state potential determined by molecular spectroscopy \cite{all03} that provided information about the interaction potential up to 2~nm internuclear separations. 

The observed levels with binding energies between 1.6~MHz and 8.2 GHz with respect to the $^1S_0 + {^1S_0}$ asymptote can be used to precisely determine the long range behavior of the interaction potential and to extract the s-wave scattering length $a$ that plays a key role for the elastic scattering properties of $^{40}$Ca at ultralow temperatures \cite{kra09}. 
In a complementary measurement we have determined the molecular dipole matrix element for one transition between vibrational states in the electronically excited and the ground state potential from the Autler-Townes splitting created by coupling of these states with a resonant laser field. 

\section{Experimental Setup}

Samples of bosonic $^{40}$Ca atoms are prepared at ultralow temperatures by two subsequent stages of magneto-optical traps (MOT) using the broad singlet ${^1 S_0}-{^1 P_1}$ and the strongly forbidden intercombination ${^1 S_0}-{^3 P_1}$ transitions \cite{deg05}. 
During the cooling phases the atoms at a temperature of 10~$\mu$K are accumulating in a crossed dipole trap \cite{kra09}. 
Following the MOT phases the temperature of the atoms in the dipole trap is further reduced by forced evaporative cooling, ramping down the dipole trap depth. 
About $2 \times 10^5$ atoms remain at a final temperature of $T \approx 1 \; \mathrm{\mu} \mathrm{K}$ and a maximum density of $\rho \approx 10^{19} \; \mathrm{m}^{-3}$ in the crossing region of the dipole trap beams for the experiments described here.
Two PA lasers precisely tunable in the range of up to $-1.4~\mathrm{GHz}$ and $-40~\mathrm{GHz}$ relative to the atomic resonance irradiate the atomic cloud. 
Both lasers have a narrow linewidth $\Delta \nu \sim 1 \;$Hz realized by an offset phase-lock to an extended cavity diode laser, which is stabilized by the Pound-Drever-Hall method \cite{dre83} to a high-finesse ULE cavity \cite{naz08}.  
Thus, the difference frequency between the PA lasers inherits the narrow linewidth, enabling ultra-high resolution two-color PA spectroscopy.   

The output power of both PA lasers is amplified by a system of injection-locked slave lasers. 
The spectral purity of the difference frequency between both PA lasers was determined from an independent beat of both lasers, which indicates about 90 \% power in a sub-Hertz-linewidth coherent peak.
The remaining power is contained in the servo peaks of the phase lock within a bandwidth of 250 kHz.
Both PA light sources are coupled to the same single-mode optical fiber to ensure that they irradiate the atomic cloud with a power of up to 20~mW from the same direction, suppressing Doppler-broadening in the recorded spectra. 
Both beams are linearly polarized parallel with respect to an applied magnetic field of $B = 285~\mathrm{\mu T}$ and focused to $1/e^2$ beam waist radii $\omega_0 = 50~\mathrm{\mu m}$ at the location of the atom cloud. 
We are using PA irradiation times of up to $200~\mathrm{ms}$ and maximum intensities of $I_1= 250~\mathrm{W cm}^{-2}$ and $I_2=100 \; \mathrm{W cm}^{-2}$ for the lasers driving the free-bound and the bound-bound molecular transition, respectively.
Depending on the measurement configuration one of the PA lasers is scanned and the induced loss of atoms from the dipole trap is determined from absorption images using the ${^1 S_0} - {^1 P_1}$ singlet transition.

\section{Signal modeling}
\label{signalmodel}
\subsection*{Theoretical lineshape}

The PA-spectra are evaluated using the theoretical model developed by Bohn and Julienne \cite{boh96,boh99}. 
It considers a pair of colliding atoms with a kinetic energy $\epsilon $ of the relative motion in the presence of two light fields with frequency $\omega_1$ and $\omega_2$, respectively (see Fig.\ \ref{fig:schema}). In the following all frequencies are expressed in angular frequencies if not noted otherwise.

The light field with frequency $\omega_1$ is tuned closely to a molecular transition frequency which couples an excited rovibrational state $\left|1\right>$ to the scattering state $\left|\epsilon\right>$. 
The detuning from this intermediate state is 
$\Delta_1$, 
i.e. for $\Delta_1 = 0$ the light field is in resonance with the free-bound transition for $\epsilon  = 0$ (see Fig.\ \ref{fig:schema}).

The second laser frequency $\omega_2$ drives the bound-bound transition between states $\left|1\right>$ and $\left|2\right>$.  
The two color-detuning with respect to the ground state bound level located $\hbar \Delta_{\mathrm{b}}<0$ below the ground state asymptote is given by
$ \Delta_2 =  \Delta_{\mathrm{b}} - (\omega_1 - \omega_2)$.  
The two-color resonance at $\epsilon  = 0$ corresponds to $ \Delta_2 = 0 $. 

Generally the lifetime of the bound state $\left|2\right>$ is much longer than the one of the intermediate state. 
Thus on the two-color resonance, the main loss mechanism proceeds through excitation to the intermediate state $\left|1\right>$, from where the photoassociated molecules decay spontaneously and produce atoms with high enough energy to escape from the dipole trap.

For samples at $\mu$K temperatures only s-wave scattering is relevant, and the probability for a two-color PA loss from the input scattering channel $\left|\epsilon\right>$ to the excited state $\left|1\right>$ is given by the squared scattering matrix element $\left|S_{\epsilon 1}\right|^2$ \cite{boh99} \footnote{Note that the Rabi frequency $\Omega_{12}$ differs from the formula given in \cite{boh99}. Our definition follows the convention in \cite{coh92} representing the frequency for transferring amplitude between the two bound states.}:
\begin{equation}
	\left|S_{\epsilon 1}\right|^2 = 
	\frac
	{\gamma_1 \Gamma_\mathrm{stim} {\Delta_2^\prime}^2}
	{\left( \Delta_1^\prime  \Delta_2^\prime - \Omega^2_{12}/4 \right)^2 
	+ \left( (\gamma_1 + \Gamma_\mathrm{stim})\Delta_2^\prime / {2}\right)^2 }
\label{eq:lshape}
\end{equation}
with
$\Delta_2^\prime =  \Delta_2 - \epsilon/\hbar $
and
$\Delta_1^\prime =  \Delta_1 - \epsilon/\hbar + \delta_1 $.
Here $\gamma_1$ denotes the total decay rate of the upper molecular state, including not radiative decay channels, 
$\hbar \delta_1$ corresponds to the light shift  of the molecular level $\left|1\right>$ induced by laser 1. 
The coupling by the laser fields is described as time-dependent interaction 
$- \textbf{d}(R) \cdot \textbf{E}_i (t) = V^{(i)}_\mathrm{opt} \cos(\omega_{i}t)$,
where the constant optical potential $V^{(i)}_\mathrm{opt}$ denotes the amplitude of the harmonic perturbation, 
$\textbf{d}(R)$ the molecular transition dipole operator, 
and $\textbf{E}_i(t) = E_{i} \cos (\omega_i t)$ the electric field of laser $(i)$  
with amplitude $E_i = \sqrt{2 I_i/\epsilon_0 c}$. 
Hence we obtain the molecular Rabi frequency
\begin{equation}
\Omega_{12} = \frac{1}{\hbar} \left < 1 | V^{(2)}_{\mathrm{opt}}|2 \right> 
\end{equation}
for a transition between states  $\left|1\right>$ and $\left|2\right>$, which will be explained in greater detail in Sec. \ref{sec:ats}.
The free-bound excitation by laser 1 from the energy-normalized continuum state $\left|\epsilon\right>$ to state $\left|1\right>$ according to Fermi's golden rule the harmonic perturbation induces a stimulated rate \cite{coh92}:
\begin{equation}
\Gamma_{\mathrm{stim}} = 
\frac{\pi}{2 \hbar}
\left| \left<1|V^{(1)}_\mathrm{opt}|\epsilon \right>\right|^2.
\label{eqn:stim}
\end{equation}
Under the assumption that $\mathbf{d}(R)$ is only weakly depending on the internuclear separation $R$ at long range, the total radiative decay rate of the atom pair considered here is twice the atomic decay rate
$\gamma_{\mathrm{atom}}$, and the stimulated rate can be expressed as \cite{ciu06}:
\begin{eqnarray}
\Gamma_{\mathrm{stim}}=
\gamma_{\mathrm{atom}}  \frac{3 I_1 c^2 \pi^2}{\omega^3} f_{\mathrm{ROT}} f_{\mathrm{FCD}},
\label{eqn:FCD}
\end{eqnarray}
where $f_{\mathrm{ROT}}$ contains the factor of 2 between atom pair and atomic decay rate, the polarisation dependence and the H\"onl-London factor. $f_{\mathrm{FCD}}  = \left| \left<1|\epsilon \right>_{\mathrm{rad}}\right|^2$ 
is the Franck-Condon density between the energy-normalized scattering state $\left|\epsilon \right>$ and the bound state $\left|1\right>$, where the subscript indicates that only the radial component is taken into account. $\omega$ is the atomic transition frequency and $I_1$ the laser intensity.
At low energy $\epsilon$ the Wigner threshold law applies, which allows to express $\Gamma_{\mathrm{stim}}$ as:
\begin{equation}
\Gamma_\mathrm{stim} = 2  k l_\mathrm{opt} \gamma_1
\label{eqn:stim1}
\end{equation}
with an energy-independent optical length $l_\mathrm{opt}(I_1)$, the scattering wavenumber 
$ k=(2 \mu \epsilon)^{1/2} / \hbar$, and the reduced mass $\mu$ of the calcium dimer. 
%
%
The inelastic loss coefficient $K$ for a thermal sample at temperature $T$ is the thermal average
\begin{equation}
K = 
\frac{1}{hQ_{\mathrm{T}}} 
\int_0 ^\infty  e^{-\epsilon / k_\mathrm{B} T} \left|S_{\epsilon 1}\right|^2 d\epsilon ,
\label{eq:thermal_av}
\end{equation}
where $Q_\mathrm{T} = ({2 \pi \mu k_\mathrm{B} T}/{h^2})^{3/2}$ denotes the partition function for free particles.
%
The total loss of atoms, assuming that all molecules photoassociated to state $\left|1\right>$ will decay to unbound atom pairs with enough energy to leave the dipole trap, can be described by the differential equation:
\begin{equation}
\dot{\rho} = -\alpha \rho - 2 K \rho^2,
\label{eq:dynamics}
\end{equation}
where $\rho$ denotes the local density of atoms inside the trap, $\alpha$ the loss coefficient from collisions with background gas and $K$ the loss coefficient for PA-induced losses. 
In our experiment only short irradiation times $\tau$ compared to the unperturbed trap lifetime $\alpha^{-1}$ were used, thus inelastic scattering with background gas can be neglected i.e. $\alpha \approx 0$.
In this case, the differential equation Eq.\ \ref{eq:dynamics} can be integrated across the trapping volume to give the atom number $N(t)$ assuming that the initial thermal Gaussian distribution remains Gaussian throughout the PA process. The differential equation for the total number of atoms can then be solved:
\begin{equation}
N(\tau) = \frac{N_0}{1+2 \tau N_0 K/( \sqrt{8} V_\mathrm{eff})},
\label{eq:dgl}
\end{equation}
with an effective trap volume 
\begin{equation*}
V_\mathrm{eff}= \frac{1}{\omega_x \omega_y \omega_z} \left( \frac{{2 \pi k_\mathrm{B} T}}{m_{\mathrm{Ca}}}\right)^{3/2}
\end{equation*}
and trap frequencies $\omega_i$~($i=x,y,z$).
\begin{figure}
\includegraphics[width=\columnwidth]{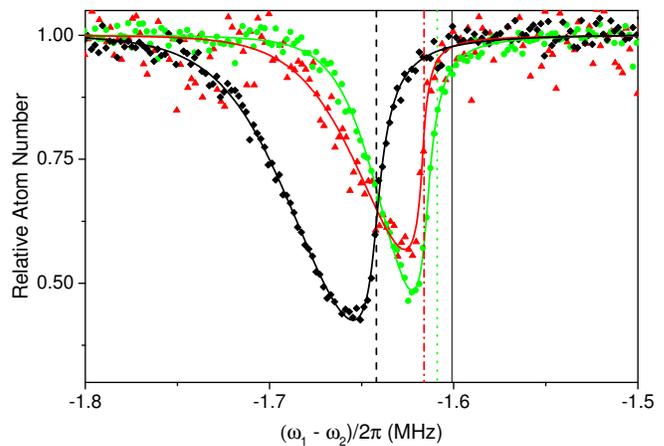}
\caption{Atom loss from the crossed dipole trap due to two-color PA spectroscopy as a function of the difference of the two laser frequencies for the most weakly bound molecular state in the ground state potential \Xpot $v_2~=~40;\;J_2~=~0$
using $v_1~=~-1;\;J_1~=~1$ in $\mathrm{c}(0^+_u)$ as the intermediate level. 
Spectra for three different PA-intensities are shown together with their respective fit curves and their derived resonance position indicated by vertical lines. 
The green dots and the vertical dotted line correspond to $I_1 =  45~ \mathrm{Wcm}^{-2}$ and $I_2 = 18~ \mathrm{Wcm}^{-2}$, black diamonds and the vertical dashed line to $I_1 = 118~ \mathrm{Wcm}^{-2}$ and $I_2 = 18~ \mathrm{Wcm}^{-2}$, red triangles and the dashed-dotted line to $I_1 =  45~ \mathrm{Wcm}^{-2}$ and $I_2 = 14 ~\mathrm{Wcm}^{-2}$. The solid line at -1.601 MHz indicates the unperturbed binding energy.}
\label{fig:2colorPA}
\end{figure}
In the so-called Raman configuration for two-color PA the laser driving the free-bound transition is far detuned from the intermediate molecular level i.e. $\Delta_1 \gg (\gamma_1,\Gamma_{\mathrm{stim}},\Omega_{12})$.
Under these conditions a maximum of the matrix element $|S_{\epsilon 1}|^2$ is located at $\Delta_2^\prime = \frac{\Omega_{12}^2}{\Delta_1^\prime}$. Expressing Eq.\ \ref{eq:lshape} in terms of the detuning from this maximum, i.e. $\Delta_2^{\prime\prime} = \Delta_2^{\prime} - \frac{\Omega_{12}^2}{\Delta_1^\prime}$, and assuming to stay in the vicinity of the maximum far away from the minimum at $\Delta^{\prime}_2 = 0$ ($|\Delta_2^{\prime\prime}| \ll |\frac{\Omega_{12}^2}{\Delta_1^\prime}|$), the individual lineshape for a fixed collision energy can be approximated by a Lorentzian:
\begin{equation}
\left|S_{\epsilon 1}(\Delta_2)\right|^2 \approx \frac{A}{(\Delta_2 - \epsilon/\hbar + \delta_{\mathrm{shift}})^2 + (\Gamma_L/2)^2}
\end{equation}
with parameters $A,\; \delta_{\mathrm{shift}},\; \Gamma_L$ which are related to the experimental parameters:
\begin{eqnarray}
\Gamma_{\mathrm{L}} &=&  \frac
	{\Omega^2_{12} (\Gamma_\mathrm{stim} + \gamma_1)}
	{4(\Delta_1 -\epsilon/\hbar + \delta_1)^2},
	\label{eqn:lorwidth}  \\
\delta_{\mathrm{shift}} &=&  \frac
	{\Omega^2_{12}}
	{4(\Delta_1 -\epsilon/\hbar + \delta_1) }, 
	\label{eq:shift} \\
A &=& \frac
	{\Gamma_{\mathrm{stim}} \gamma_1 \Omega_{12}^4}
	{16(\Delta_1 -\epsilon/\hbar + \delta_1)^4}.
	\label{eq:amplitude}
\end{eqnarray}
In the Raman configuration as used in our setup the deviation between the approximation and the true individual line shape is only marginal and thus can be neglected when performing the thermal averaging over collision energies $\epsilon $. When applying Eqs. \ref{eqn:lorwidth}, \ref{eq:shift}, \ref{eq:amplitude} we neglect the kinetic energy $\epsilon$ in the denominator, because its contribution is small compared to $\Delta_1+\delta_1$. This approximation allows an efficient modeling of the thermally broadened measured spectra (see Fig.\ \ref{fig:2colorPA}) as applied in section \ref{ground}.
Similar to one-color PA \cite{jon99}, the fit approximates the integral of the thermal average (Eq.\ \ref{eq:thermal_av}) by a sum of lines evenly spaced by $\delta \epsilon \approx \hbar \Gamma_\mathrm{L}/3$ 
\begin{equation}
K \approx  \sum_{n=0}^N e^{- n\cdot \, \delta \epsilon / k_\mathrm{B} T} \left|S_{\epsilon 1}\right|^2\, \delta \epsilon .
\label{eq:K}
\end{equation}
The number $N$ of summation intervals has to be large enough to cover the entire line profile and is mostly depending on the temperature of the sample.

\section{Measurements Results}
\label{results}
\subsection*{Ground state binding energies}
\label{ground}

We have observed two-color PA spectra of four weakly bound molecular states $\left| 2 \right>$ in the potential \Xpot via intermediate bound states $\left| 1 \right>$ of the two excited states denoted
$\mathrm{a} ^3 \Sigma^+_{\mathrm{u}}$ and $\mathrm{c} ^3 \Pi_{\mathrm{u}}$ in Hund's coupling case (a). Near the \exasym asymptote the spin-orbit interaction becomes dominant and the adiabatic potentials are more accurately described by Hund's case (c) potentials $\mathrm{c}0^+_{\mathrm{u}}$ and $(\mathrm{a,c})1_{\mathrm{u}}$ being a strong mixture of the two case (a) states (Fig.\ \ref{fig:schema}). In the experiment we have used $v_1 = -1, J_1 = 1, \Omega = 1$ and $v_1 = -1, J_1 = 1, \Omega = 0$ \cite{Kah14} as intermediate states from which laser $1$ was detuned by $\Delta_1 / 2 \pi \approx \pm1~\mathrm{MHz}$. Here $J$ denotes the total angular momentum of the respective state.
The PA lines were then fitted using the procedure described before to assess the resonance frequency at zero-collision energy i.e. $\epsilon = 0$. 

\begin{figure}
\includegraphics[width=0.97\columnwidth]{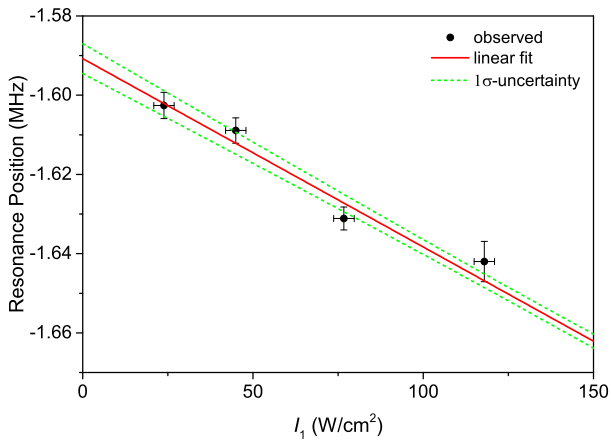}
\caption{Measured resonance position of $v = 40$, $J = 0$ depending on the free-bound PA laser intensity $I_1$. 
The solid red line shows a linear extrapolation to zero intensity while the green doted lines give the 1$\sigma$-uncertainty interval. Measurements were taken at $I_2~=~18.3~\mathrm{Wcm}^{-2}$, $I_{\mathrm{DT}}~=~ 44.2~ \mathrm{Wcm}^{-2}$ }.
	\label{fig:L1shift}
\end{figure}

\begin{figure}
\includegraphics[width=0.97\columnwidth]{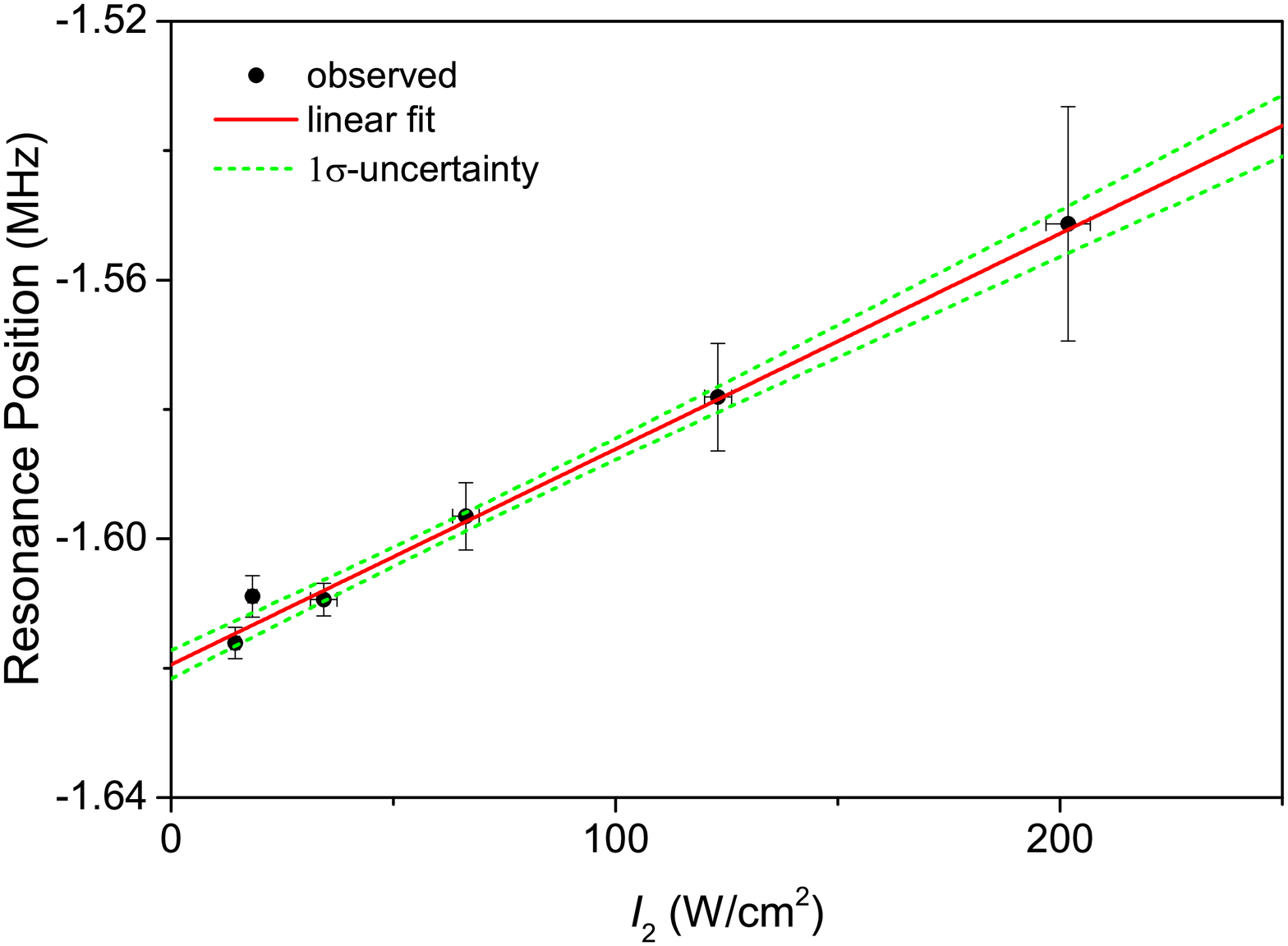}
\caption{Measured resonance position of $v = 40$, $J = 0$ depending on the bound-bound PA laser intensity $I_2$. 
The solid red line shows a linear extrapolation to zero intensity while the green doted lines give the 1$\sigma$-uncertainty interval. Measurements were taken at $I_1~=~45.0~\mathrm{Wcm}^{-2}$, $I_{\mathrm{DT}}~=~ 44.2~ \mathrm{Wcm}^{-2}$ }.
	\label{fig:L2shift}
\end{figure}

\begin{figure}
\includegraphics[width=0.97\columnwidth]{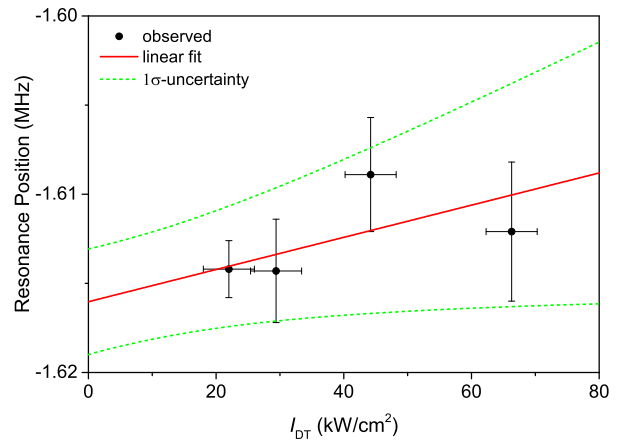}
\caption{Measured resonance position of $v = 40$, $J = 0$ depending on the dipole trap laser intensity $I_{\mathrm{DT}}$. 
The solid red line shows a linear extrapolation to zero intensity while the green doted lines give the 1$\sigma$-uncertainty interval. Measurements were taken at $I_1~=~45.0~\mathrm{Wcm}^{-2}$, $I_2~=~ 18.3~ \mathrm{Wcm}^{-2}$ }.
	\label{fig:LDTshift}
\end{figure}

Three spectra of the most weakly bound state $v_2=40;\; J_2=0$ in X $^1\Sigma^+$ for different experimental parameters and their respective fitted lineshapes are shown in Fig.\ \ref{fig:2colorPA}. Their characteristics shall be discussed in the following. The level $v_1 = -1;\; \Omega = 0$  has been used as the intermediate level in the Raman configuration PA with the large detuning $\Delta_1 / 2 \pi \approx +1~\mathrm{MHz}$. 
All spectra were taken at a temperature of $T = 1.15~\mu \mathrm{K}$ determined from time of flight measurements. 
They all show the typical asymmetric thermal broadening that is a characteristic feature of photoassociation spectra when the mean kinetic energy is large compared to the width of the excited level. If we used the energy dependence of $\Gamma_\mathrm{L}$ in Eq.\ \ref{eqn:lorwidth} through $\Gamma_{\mathrm{stim}}$ according to Eq.\ \ref{eqn:stim1} and assumed $\gamma_1 = 2 \gamma_{\mathrm{atom}}$ during the fitting procedure we obtained nonphysical values for the set of parameters $\{ T, \Omega_{12}, l_{\mathrm{opt}} \}$. This behavior points towards a larger effective decay rate $\gamma_1$. A similar effect of enlarged molecular decay rates has been reported for PA spectra of $^{88}$Sr \cite{zel06}. To compensate for this we use an energy-independent effective linewidth $\Gamma_\mathrm{L}$ for a given experimental condition defined by the intensities of laser 1 and 2, detuning $\Delta_1$ and temperature T. For the fit of the profiles according to Eq.\ \ref{eq:K} we represent the squared matrix element $\left|S_{\epsilon 1}\right|^2$:
\begin{eqnarray}
\left|S_{\epsilon 1}(\omega_1 - \omega_2)\right|^2 \approx \nonumber \\
\frac{A^{\mathrm{eff}}\sqrt{\epsilon}}{(-(\omega_1 - \omega_2) - \epsilon/\hbar - \Delta_b^{\mathrm{eff}})^2 + (\Gamma_L/2)^2}
\end{eqnarray}
with effective fit parameters $\Delta_b^{\mathrm{eff}}$ that contains light shifts induced by the spectroscopy and trapping lasers (compare Figs.\ \ref{fig:L1shift}, \ref{fig:L2shift}, \ref{fig:LDTshift}) and $A^{\mathrm{eff}} = A / \sqrt{\epsilon}$ to account for the energy dependence of the stimulated rate in $A$ .
Increasing the free-bound intensity $I_1$ from the green (circles) to the black (diamonds) curves in Fig.\ \ref{fig:2colorPA} leads to a considerable light shift of the resonance position.
The linewidth $\Gamma_{\mathrm{L}}$ increases from $~6.0 ~\mathrm{kHz}$ to $~9.2 ~\mathrm{kHz}$ while the
temperature derived by the fit remains nearly constant about 20 \% below the value determined from the time of flight measurement. 
This behavior is connected to the change in the stimulated rate in Eq.\ \ref{eqn:FCD} or Eq.\ \ref{eqn:stim1} and to an increased light shift $\delta_1$, leading to a broader individual linewidth. 

In the same way a decrease of bound-bound laser intensity between the green (dots) and red (triangles) data points leads to a decrease of $\Gamma_{\mathrm{L}}$ from $6.0~\mathrm{kHz}$ to $4.3~\mathrm{kHz}$. This is consistent with a reduced molecular Rabi frequency $\Omega_{12}$ that also shifts the resonance position according to Eq.\ \ref{eq:shift}. 
The fit correctly reproduces the theoretical expectations regarding individual PA linewidth and PA rates as a function of increasing PA intesities $I_1,I_2$. The sample temperature determined by the fit is slightly underestimated. 

Experimental noise near the steep edge in the spectra can have enormous impact on some of the fit parameters.
Fortunately, the resonance position determined by the fit is very robust against changes in $\Gamma_\mathrm{L}, \Delta_b^{\mathrm{eff}}, A^{\mathrm{eff}}, T$ and is mostly determined by the steep edge, that is only a few kHz wide (see Fig.\ \ref{fig:2colorPA}). 
Thus the zero-energy resonance position of the molecular level can be determined with high reliability.

In addition to the thermal shift, the observed line is also shifted due to the ac Stark effects from the two PA lasers and from the trapping laser and should be contained in the parameter $\Delta_b^{\mathrm{eff}}$. These shifts were determined independently by extrapolation to zero laser intensity from a series of measurements for different intensities as shown in Figs. \ref{fig:L1shift}, \ref{fig:L2shift}, \ref{fig:LDTshift} to reduce the model dependence from the representation of $\left|S_{\epsilon 1}\right|^2$ and to take into account the effect by the trapping laser when deriving the desired binding energies from resonance positions.
An overview of all considered corrections on the final resonance position and their respective uncertainties is given in Tab.\ \ref{tab:uncertainty}. 
An additional uncertainty contribution of the ac Stark shift by laser 2 arises according to Eq.\ \ref{eq:shift} by the detuning of the intermediate level. Multiple measurements with detuning $\Delta_1/2 \pi$ in the range $\pm(400-1000)~\mathrm{kHz}$ indicate an uncertainty of 1~kHz for the resonance position.

As mentioned above, the fit routine underestimates the temperature of the atomic sample by about 20 \% compared to the temperature from time-of-flight measurements. 
We account for this deviation by an uncertainty of the derived binding energy of $ 0.1 \cdot \epsilon/\hbar \approx 2~$kHz (see Tab.\ \ref{tab:corrections}).
 
We estimate the combined uncertainty of the aforementioned sources combined with other sources of technical nature to be 3~kHz.
Among these are small nonlinearities of the atom number estimation from the absorption imaging, variations of dipole trap power and the ramp procedure for the evaporation cooling, residual contributions of unwanted frequencies stemming from the offset-locking of the spectroscopy lasers and residual magnetic fields. 

The unperturbed binding energies  $\hbar \Delta^{\rm exp}_{\rm b}$ including all corrections and uncertainties are shown in Tab.\ \ref{tab:corrections}. 
\begin{table*}[t]
\begin{center}
 \begin{tabular}{|c|c||c|c||c|c|c||c|}
\hline
  bound level & $v=38\; J=2$ & \multicolumn{2}{|c||}{$v = 39\; J = 0$} & \multicolumn{3}{|c||}{$v = 39\; J = 2$} & $v = 40 \; J = 0$ \\ 
\hline
intermediate state  & $v_1=-1 \Omega=0 $ & \multicolumn{2}{|c||}{$ v_1=-1 \; \Omega=1 $}  & \multicolumn{3}{|c||}{$ v_1=-1 \; \Omega=1 $} &$ v_1=-1\; \Omega=0$\\
detuning $\Delta_1 / (2\pi)$ & -1\,101(10) & -1\,001(10) & 1\,016(10) & -1\,001(10) & -924(10) & +900(10) & +978(10)	\\
\hline
\hline
ac Stark free-bound laser	& 1(2) & 0(10)	& -1(6)		& 0(3)	& 0(5) & -24(24)	& 22(4)\\ 
ac Stark bound-bound laser	& 4(2) & 30(5) 	& -9(5)		& 9(3)	& 8(3) & -10(2) 	& -6(2)\\
ac Stark dipole trap 	& 1(3) & -17(17)& -4(4) 	& 0(4)	& 0(3) & 0(3) 	& -3(3)\\
\hline
extrapolated position & -8\,218\,897(4) & -1\,387\,457(20) & -1\,387\,439(9)& -1\,005\,369(6) & -1\,005\,361(7) & -1\,005\,379(24) & -1\,601(6)\\
\hline 
\hline 
weighted average	& -8\,218\,897(4) & \multicolumn{2}{|c||}{-1\,387\,442(8)} & \multicolumn{3}{|c||}{-1\,005\,366(5)} & -1\,601(6)	\\
\hline
systematic uncertainty	& 0(3) & \multicolumn{2}{|c||}{0(3)} & \multicolumn{3}{|c||}{0(3)} & 0(3) \\
\hline
\hline
unperturbed energy $\Delta^{\rm exp}_{\rm b} / (2\pi)$	& -8\,218\,897(5) & \multicolumn{2}{|c||}{-1\,387\,442(9)} & \multicolumn{3}{|c||}{-1\,005\,366(6)} & -1\,601(7)	\\
\hline
calc. energy $\Delta_{\mathrm{calc}} / (2\pi)$ & -8\,218\,896 & \multicolumn{2}{|c||}{-1\,387\,437} & \multicolumn{3}{|c||}{-1\,005\,372} & -1\,599	\\
\hline
\end{tabular}
\caption{
The binding energies $\hbar \Delta^{\rm exp}_{\rm b}$ with corrections and corresponding uncertainties. 
All values are given in kHz.}
\label{tab:corrections}
\label{tab:uncertainty}
\end{center}
\end{table*}
\section{Molecular Potentials}
 
Molecular potentials are needed for the prediction of binding energies, calculating Franck-Condon densities (FCD) in Eq.\ \ref{eqn:FCD} , and modeling bound-bound transitions with appropriate Franck-Condon factors (FCF). 

The excited state potentials at the asymptote \exasym were determined in \cite{Kah14} and the vibrational levels are represented by a multi-component wave function, 
which we will describe most conveniently for the calculation of the electric dipole transition by the Hund's case (e) basis $\left|{^1S_0} + {^3P_j},l,J\right>$. 
The first part of the basis vector describes the relevant atom pair asymptote with the total atomic angular momentum $j$, $l$ the angular momentum of the pair rotation and $J$ the total angular momentum. 
Because the starting level in the present experiment is always the s-wave continuum, corresponding to $J=0$ and parity +, we have only to consider the excited levels $J=1$, parity -,  thus only the three possible basis states
\begin{align}
&\left| ^1S_0 + {^3P_1},0,1\right>, \\
&\left| ^1S_0 + {^3P_1},2,1\right>, \label{comp1} \\ 
\text{and} &\left| ^1S_0 + {^3P_2},2,1\right>. \label{comp2}
\end{align}
Because of the electric dipole selection rule $\Delta l=0$ from the calculated eigenstates for J=1 only the component 
$\left|^1S_0+{^3P_1},0,1\right>$ is needed for the calculation of the Franck-Condon densities with the continuum $ \left |^1S_0+{^1S_0},0,0 \right >$. The first or second component(Eq. \ref{comp1}, \ref{comp2}, respectively) were used for the Franck-Condon factors with the bound levels 
$\left|^1S_0+{^1S_0},0,0\right>$ ($J = 1 \rightarrow 0$ transition) or $\left|^1S_0+{^1S_0},2,2\right>$ ($J = 1 \rightarrow 2$ transition), respectively.

The molecular ground state potential \Xpot is based on spectroscopic measurements by \cite{all03} for the short range part and on the new data for the long range branch as described by the conventional power expansion in $1/R$
\begin{equation}
V_{X}= -f_6(R)\frac{C_{6}}{R^6} -\frac{C_{8}}{R^8}-\frac{C_{10}}{R^{10}}.  
\label{eq:potential}
\end{equation}
By the function $f_6(R)$ we apply the retardation correction as calculated by Moszynski \textit{et al.} \cite{mos03}, which turned out to be essential for describing the asymptotic levels within their experimental uncertainty. This correction is only significant to the first term in the equation due to the very long range nature of the van der Waals interaction. An additional exchange term is regularly used in the mathematical representation of the long range potential. Such term is not needed for the description in our case, due to the van der Waals term being large compared to a possible exchange energy in the range $R > 1.1~\mathrm{nm}$ in which Eq.\ \ref{eq:potential} is applied.
\begin{table}[h]
\centering
\begin{tabular}{|c||c|c|c|}
\hline
ref. &  $C_6$   &  $C_8$ &  $C_{10}$\\
\hline
 &  $10^{7}$ \wn \AA$^6$  & $10^{8}$ \wn \AA$^8$ & $10^{9}$ \wn \AA$^{10}$ \\
\hline
  \cite{all03} &  1.0023 & 3.808 & 5.06\\
  \cite{ciu04} & 1.003  &  & \\
  \cite{por06a} & 1.022 & 3.010  & 8.057	\\
  \cite{mit08} & 1.055 & 3.060 & 8.344	 \\
	this work & 1.0348 & 2.997 &10.88    \\  
\hline
\end{tabular}
\caption{Comparison of long range parameters at the asymptote  ${^1S_0}+{^1S_0}$ derived in this work with experimental results from Allard \textit{et al.} \cite{all03} and theoretical results from Ciury{\l}o \textit{et al.} \cite{ciu04}, Porsev \textit{et al.} \cite{por06a} and  Mitroy and Zhang \cite{mit08}. No error limits are given for the individual long range parameters from this work because of significant correlation between these parameters. The full set of long range parameters should always be applied for model calculations.}
\label{tab:Ci}
\end{table}
%
\begin{table}[h]
\centering
\begin{tabular}{|c||c|}
\hline
method &  $a/a_{0}$   \\
\hline
  molecular spectroscopy \cite{all03} & 200 - 800   \\
	$^1S_0 + ^1P_1$ photoassociation \cite{vog07} & 340 - 700 \\
	BEC mean field \cite{kra09} & $\approx 440$ \\
	this work &  308.5 (50) \\
\hline
\end{tabular}

\caption{${^1S_0}+{^1S_0}$ scattering length $a$ of $^{40}\mathrm{Ca}$ in comparison with experimental results, $a_0\approx 53~\mathrm{pm}$ denotes the Bohr radius}
\label{tab:scatter}
\end{table}

In a least squares fit of all known data for the ground state (in total 3586 data points) the long range parameters were varied including as additional condition the theoretical coefficients $C_i$ for $i=6, 8, 10$ \cite{por02,por06a,mit08,ciu04} applying their estimated uncertainties from the calculation as weights in the fit. 
The fit reveals in total a reduced $\chi^2 = 0.58$, 
a good fit should result to values close to one, if the applied uncertainties are well justified and the theoretical model is appropriate.
Thus we obtain a very satisfactory result. 
Trying combinations of $C_6$, $C_8$ and $C_{10}$, we see the statistically allowed spread of the coefficients with their correlation and conclude that the accuracy of the derived $C_6$ values is about 3\% and that of $C_8$ is about 15\%, but the total long-range function will be much better in the range $R > 1.6$~nm because of the correlation between the parameters. The fit also includes the data from Fourier transform spectroscopy \cite{all03}, which reach the long range region by high vibrational states. Thus the fit also slightly changes the inner part of the potential.

In detail, the spectroscopic data from \cite{all03} are represented within their respective experimental uncertainty, and the binding energies of the asymptotic levels in the last line of Tab.\ \ref{tab:uncertainty} are within their assumed uncertainties, only level v~=~39, J~=~2 touches the upper edge. 
The derived long range parameters are shown in Tab.\ \ref{tab:Ci} and are compared with earlier results from experiment and theory. 
With this improved potential representation the uncertainty of the calculated scattering length $a = 308.5(50) a_0$ decreases by more than a factor of 10 compared to former results (Tab.\ \ref{tab:scatter})
The remaining uncertainty originates to a significant part from the correlation between the long range parameters $C_6, C_8, C_{10}$. A much denser set of eigenvalues at the ground state asymptote would be required to break this correlation. 
Because the last bound state is very close to the asymptote, namely -1.6 MHz, we compared our scattering length $a$ to a simple semiclassical approximation between binding energy and scattering length \cite{gri93}
\begin{equation}
a \approx \bar{a}+\frac{\hbar}{\sqrt{2\mu E_b}},
\label{eq:scat}
\end{equation}
where $\bar{a}$ is the background scattering length, $\mu$ the molecular reduced mass and $E_b$ the binding energy. $\bar{a}$ depends on the long range form of the potential. Using only the $C_6$ value from Tab.\ \ref{tab:scatter} one calculates $\bar{a}=53.4~a_0$, and with our measured binding energy one obtains $a=291~a_0$ which comes close to the value obtained from the full potential. The difference indicates, that $E_b$ is still not small enough to accurately estimate the scattering length using this approximation.

From mass scaling we also calculate the scattering length of other natural isotopes of calcium  as shown in Tab.\ \ref{tab:scaling} using the full potential of \Xpot. 
To account for possible corrections to the assumed Born-Oppenheimer approximation we we enlarged the estimated uncertainty for the unobserved isotope combinations in Tab. IV.
\begin{table}[h]
\centering
\begin{tabular}{|c|c|}
\hline
Isotope &  $a/a_{0}$   \\
\hline
$^{40}$Ca &  308.5(50)\\ \hline
$^{42}$Ca &  297 (6)\\ \hline
$^{43}$Ca & 43.7 (10)\\ \hline
$^{44}$Ca & 399 (7)\\ \hline
$^{46}$Ca & 1970 (20) \\ \hline
$^{48}$Ca & -281 (10) \\ \hline
\end{tabular}
\caption{Scattering length $a$ of homonuclear pairs for different natural isotopes of calcium, $a_0\approx 53~\mathrm{pm}$ Bohr radius}
\label{tab:scaling}
\end{table}
Scattering lengths for other isotopes of Ca are reported \cite{dam11,zha13c} using earlier experimental results on $^{40}$Ca from our group \cite{all03, kra09}, thus these values suffer from the lower precision of those data.

The newly determined potential was applied to calculate the desired FCD and FCF (see Tab.\ \ref{tab:fcf}) for evaluation of the spectroscopic observations as detailed in section \ref{signalmodel} and the following paragraph. 

\section{Autler-Townes spectra and molecular dipole matrix element}
\label{sec:ats}

We have also used a different configuration of PA laser detunings, where the bound-bound laser was tuned to resonance, i.e. $\Delta_2 - \Delta_1 \approx 0 $, and instead the free-bound laser was scanned across the excited state resonance. Through coupling of the two bound states by the resonant light field dressed states are created that lead to two resonances: the Autler-Townes doublet \cite{aut55}. The two resonances are separated by the molecular Rabi frequency $\Omega_{12}$:
\begin{equation}
\Omega_{12} = \sqrt{f_\mathrm{ROT}} \sqrt{f_{\mathrm{FCF}}} \sqrt{\gamma_{\mathrm{atom}}} \sqrt{I_2 \frac{6 c^2 \pi}{\omega^3 \hbar}}.
\label{eq:rabi}
\end{equation}%
Here $f_\mathrm{ROT}$ is the rotational factor related to the coupling of $J_1, M_1 - J_2, M_2$ and the ratio of atomic and molecular decay rate. 
The Franck-Condon factor $f_{\mathrm{FCF}}=\left|\left<1|2\right>_{\mathrm{rad}}\right|^2$ giving a measure for the strength of molecular transitions is related to the overlap integral of the wave functions of the bound states using the selection rule for electric dipole transitions to choose the proper component of the multi-component wavefunctions.
For the transition $J= 1 \leftrightarrow 0$ the relevant component in the excited state $\left|1\right>$ is the Hund's (e) state $\left| ^1S_0 + {^3P_1},0,1\right> $ and 
for the ground state $\left|2\right>$ the component $\left| ^1S_0 + {^1S_0},0,0\right>$.
For the coupling $J=1 \leftrightarrow 2$ the relevant components are 
$ \left| ^1S_0 + {^3P_1},2,1\right> $ in the excited state 
and $ \left| ^1S_0 + {^1S_0},2,2\right> $ in the ground state. 
The factor $f_\mathrm{ROT}$ is calculated in the Hund's case (e) basis. With conventional angular momentum algebra \cite{edm57} for a transition $\left |j_1,l_1, J_1,M_1 \right>$ to $\left |j_2,l_2,J_2,M_2 \right>$ applying the decoupling of $l_{1,2}$ because the electric dipole operator does not act in the rotational space we obtain the rotational factor:
\begin{eqnarray}
f_\mathrm{ROT}=  \nonumber \\
(2J_1+1)(2J_2+1)w6j(j_1,J_1,l_1,J_2,j_2,1)^2\delta(l_1,l_2)  \nonumber \\ 
w3j(J_1,1,J_2,-M_1,q,M_2)^2 \cdot 2 \cdot (2j_1+1) ,    
\label{eq:ROT}
\end{eqnarray}
where $w3j$ and $w6j$ are the conventional Wigner-nj symbols and $q=0,\pm1$ indicates the polarization of the light field. For the present two cases with $\pi$ polarized light and $M_1=M_2=0$ the values are 2 and 4/5, respectively. 

\begin{figure}
\includegraphics[width=0.97\columnwidth]{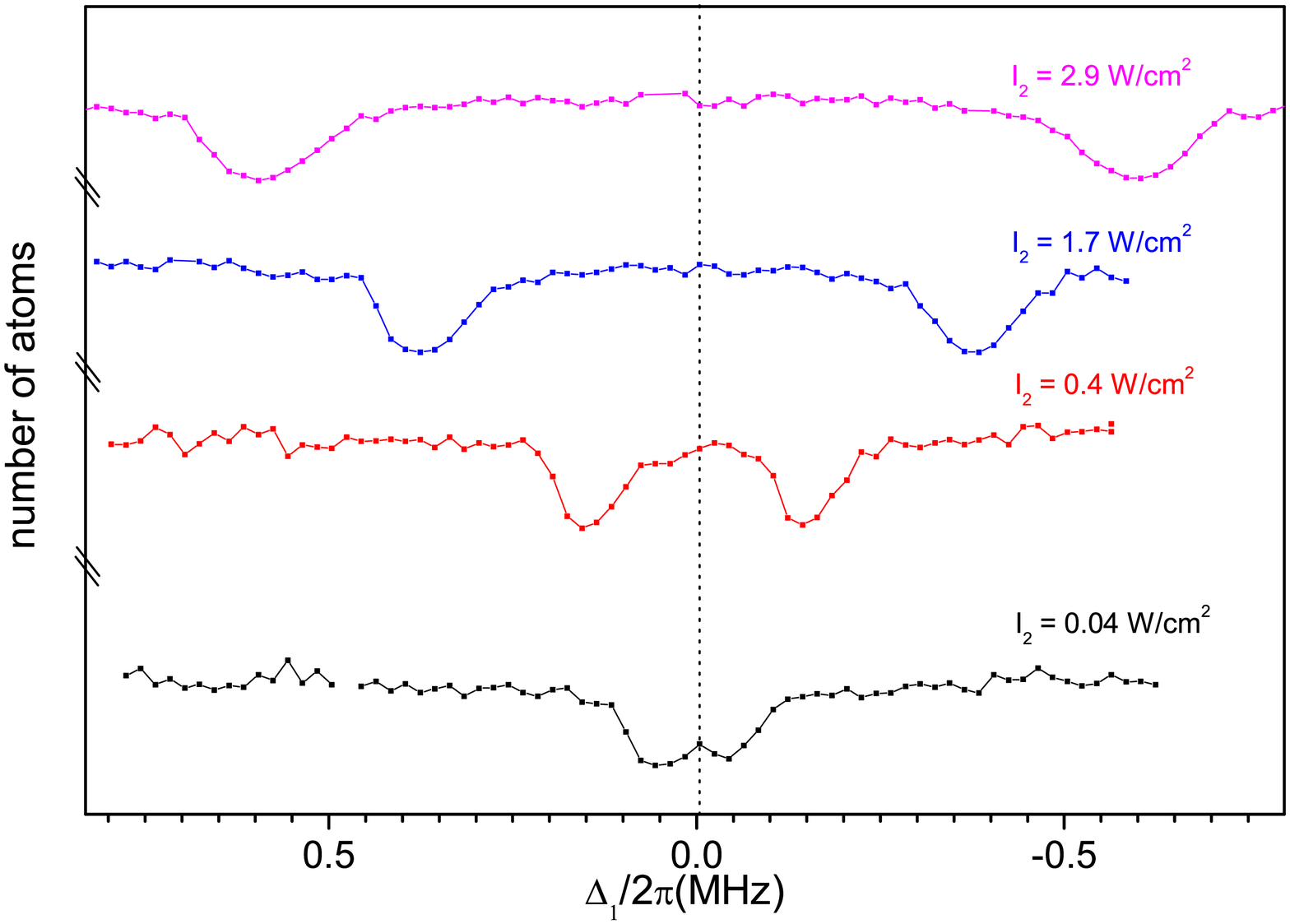}
\caption{Atomic loss spectra when scanning the free-bound PA laser across the $v_1=-1,~ \Omega = 1$ molecular resonance in Autler-Townes configuration. The bound-bound laser was set to be resonant between $v_2 = 39,~J_2=0$ and the before mentioned excited state.
Autler-Townes splitting for different bound-bound PA laser intensity $I_2$.}
\label{fig:splitting}
\end{figure}

We have measured the splitting of the Autler-Townes doublet for different intensities $I_2$ to validate the approximation that the atomic dipole moment governs the molecular transition between these long range levels. 
The recorded spectra are shown in Fig.\ \ref{fig:splitting}. 
The resonance curves show two features of almost identical amplitude indicating that the second PA-laser was indeed tuned closely to the bound-bound transition, i.e. the detuning was $\Delta_1 - \Delta_2 \approx 0$ \cite{boh99}. 

\begin{figure}
\includegraphics[width=0.97\columnwidth]{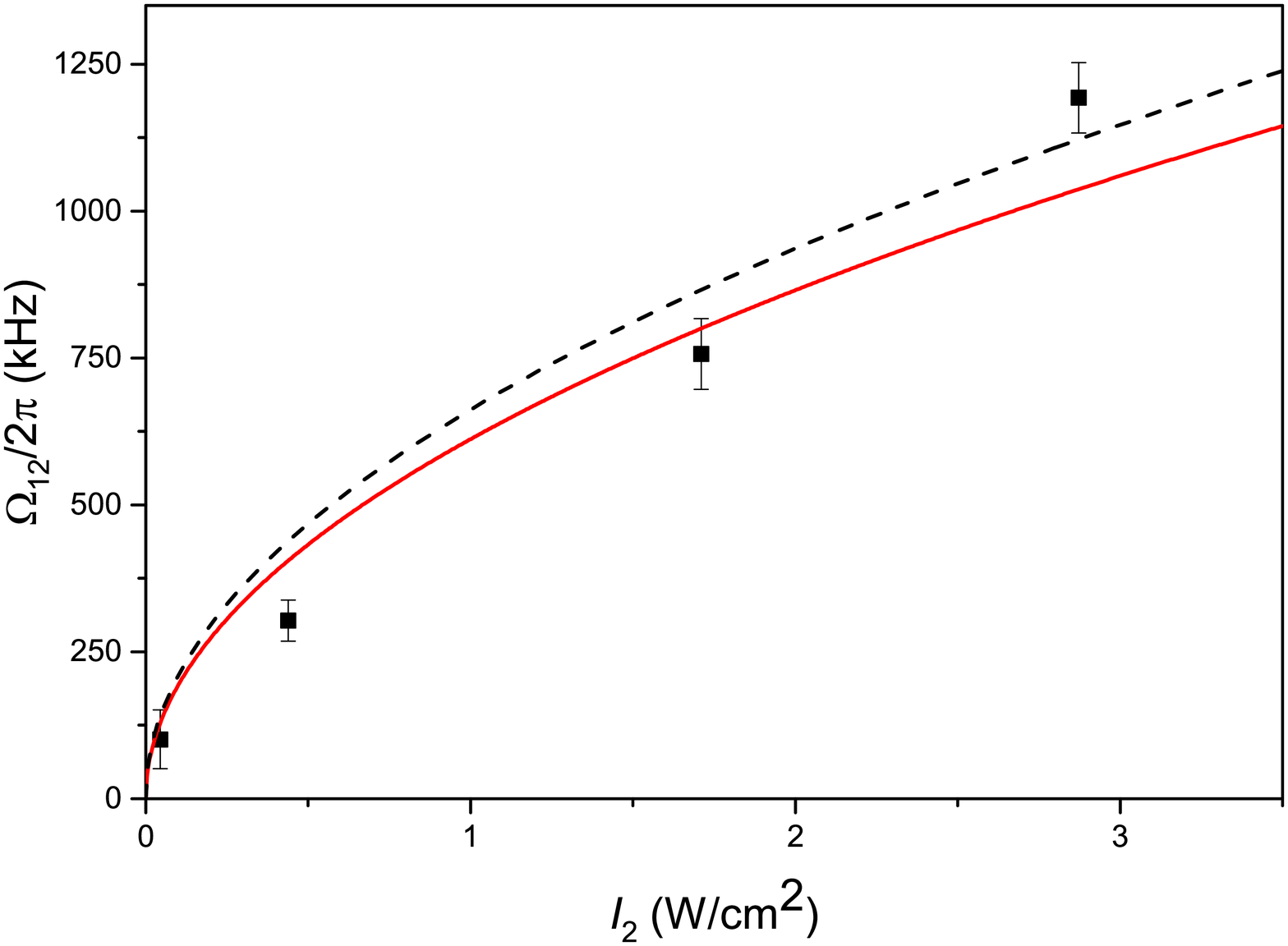}
\caption{Measured molecular Rabi frequencies $\Omega_{12}$ from Autler-Townes spectra (dots) of the $v_1=-1,\; \Omega = 1$ molecular resonance in dependence of the bound-bound laser intensity $I_2$. 
The black dashed line indicates theoretical prediction following Eq.\ \ref{eq:rabi} with $f_{\mathrm{fcf,theo}} = 0.54$ determined from molecular potentials while the red solid line is a fit to the data.}
\label{fig:Rabi}
\end{figure}

From the measured splitting we determined the Rabi frequencies for different PA laser intensities, which are plotted as a function of laser intensity $I_2$ in Fig.\ \ref{fig:Rabi}. They are consistent with the expected square-root behavior (red solid line).
From the experimental values we determined the Franck-Condon factor for a transition from the level $v_1 = -1,~J_1=1$ of the excited state state $(ac)~\Omega=1$ to the level $v_2 = 39,~J =0$. A fit of Eq.\ \ref{eq:rabi} to the data yields $f_{\mathrm{FCF,exp}} = 0.45(9)$, which is in fair agreement to the theoretical prediction $f_{\mathrm{FCF,theo}}=0.54$ (see Tab.\ \ref{tab:fcf}). The dashed line in Fig.\ \ref{fig:Rabi} shows the theoretical expectation. The small deviation could be explained by an additional uncertainty of the absolute intensity of the bound-bound PA laser at the position of the ultracold atoms due to uncertainty from the beam waist measurement and the alignment of dipole trap and focus of the PA-beam or alternatively the assumed molecular transition dipole moment deviates slightly from being $\sqrt{2}$ times the atomic dipole moment. 

The molecular potentials determined from the ground state binding energies measured in Raman configuration PA and the excited state potentials \cite{Kah14} can be used for calculating $f_{\mathrm{FCF,theo}}$ for arbitrary transitions between states close to the excited state asymptote \exasym and rovibrational states in $\mathrm{X}\,^1\Sigma^+_\mathrm{g}$. Tab.\ \ref{tab:fcf} shows a set of calculated values of $f_{\mathrm{FCF}}$ for those excited states with an appreciable Franck-Condon density $f_{\mathrm{FCD}}$ between the ground state continuum and the respective excited state, which would be desirable for easy PA spectroscopy. When calculating transition rates using the values given in Tab.\ \ref{tab:fcf} one also has to consider the additional $ f_{\mathrm{ROT}}$ for transitions to final states with $J_2 = {0,2} $, respectively. 
For calculating molecular radiation decay rates we have to sum over the polarization $q$, i.e. $\gamma_{1\rightarrow 2} = \gamma_{\mathrm{atom}} f_{\mathrm{FCF}} \, \Sigma_q f_\mathrm{ROT}$, which for the present case with atomic angular momenta 1 and 0 results in the value 2 independent of the molecular angular momentum. We obtain with values from Tab.\ \ref{tab:fcf} for the measured transition  a spontaneous decay rate from the state $v_1 = -1,~\Omega = 1~J_1=1$ to $v_2 = 39, J_2 = 0$ of $\gamma_{1\rightarrow 2} =  1.08 \cdot \gamma_{\mathrm{atom}}$ and to $v_2 = 39, J_2 = 2$ of $\gamma_{1\rightarrow 2} = 0.30 \cdot \gamma_{\mathrm{atom}}$. Tab.\ \ref{tab:fcf} gives also in the last line the remaining continuum contribution. The sum of all Franck-Condon factors of a single excited state differs slightly from one, because it only represents the part for electric dipole transitions while the spin-orbit mixing for the excited state results in a small component with total atomic angular momentum $j=2$, which will not decay by electric dipole radiation to the ground state with $j=0$.   

\begin{table}[t]
\centering
\begin{tabular}{|l|c||c|c|c|c|c|c|}
\hline
  & $\Omega$ & 0  & 1  & 0  & 1  & 0  & 1  \\ \hline
$J_2$ & $v_2 \;\backslash\; v_1 $    & -1 & -1 & -2 & -2 & -3 & -3 \\ \hline \hline
0 & 40       &  0.02  &  $ < 10^{-2}$  &  $< 10^{-3}$  &  $<10^{-2}$  &  $< 10^{-3}$  &  $<10^{-3}$  \\ \hline
2 & 39       &  0.38  &  0.15  &  0.25  &  0.04  &  0.03  &  0.02  \\ \hline
0 & 39       &  0.06  &  0.54  &  0.12  &  0.15  & 0.01   &  0.04  \\ \hline
2 & 38       &  $< 10^{-2}$  &  $< 10^{-4}$  &  0.23  &  0.12  &  0.40  &  0.13  \\ \hline
0 & 38       &  $< 10^{-2}$  &   $< 10^{-2}$ &  0.07  &  0.29  &  0.17  &  0.33     \\ \hline
2 & 37       &  $< 10^{-3}$  &  $< 10^{-6}$  &  $< 10^{-2}$  &  $<10^{-3}$  &  0.13  &  0.06  \\ \hline
0 & 37       &  $< 10^{-3}$  &  $<10^{-3}$  &  $< 10^{-2}$  &  $< 10^{-2}$  &  $0.05$  &  0.13  \\ \hline
 \multicolumn{2}{|c||}{cont.} & 0.53 & 0.30 & 0.31 & 0.39 & 0.17 & 0.27 \\ \hline
\end{tabular}
\caption{Calculated Franck-Condon factors $f_{\mathrm{FCF}}$ for transitions from bound levels ($v_1,J_1=1$) of the states $\Omega = 1_u$ and $0^+_u$ of the coupled system $\mathrm{a} ^3\Sigma^+_{\mathrm{u}}$ and $\mathrm{c} ^3\Pi_{\mathrm{u}}$ to bound levels ($v_2,J_2=0, 2$) of the ground state $X ^1\Sigma^+_{\mathrm{g}}$. The last line gives the fraction of excited molecules decaying back to continuum states near the ground state asymptote with $J = 0,2$}
\label{tab:fcf}
\end{table}
 
\section{conclusion}

This paper shows the results of our study of the ${^1S_0}~-~{^1S_0}$ asymptote of the calcium dimer by means of two-color photoassociation in Raman and Autler-Townes configuration. 
The investigation of the collisional properties of ultracold thermal calcium molecules near the dissociation limit gives access to the previously experimentally not well characterized long-range part of the ground state potential \Xpot.
Including earlier measurements of deeply bound states \cite {all03}, an improved description of the molecular potential near the dissociation asymptote was derived and also a highly improved value for the $^{40}\mathrm{Ca}$ s-wave scattering length $a = 308.5(50)\,a_0$ was calculated. The scattering length for other natural isotopes of calcium were also derived using the improved description of the ground state potential. 

With the help of a coupled channel calculation, transition moments between excited \cite{Kah14} and ground state levels have been derived and compared with experimental results from Autler-Townes measurements. The values show good agreement and confirm the validity of the theoretical description. This will allow calculations of transition paths for efficient creation of molecules in predetermined rovibrational states. According to Tab \ref{tab:fcf} for every asymptotic excited state only a few transitions to bound ground states show a substantial FCF allowing spontaneous decay to these states for use as a starting level in future experiments like coherent population transfer. For example about 54 \% of the molecules produced in the excited state $v_1 = -1,~J_1=1$ are expected to decay to the ground state $v_2 = 39,~J_2=0$.

The improved understanding of the ground state allows the calculation of an optical length $l_{\mathrm{opt}}$ for the different excited molecular states. 
For an incident intensity of $I_1 = 3.0~\mathrm{W cm^{-2}}$ we calculate a value of about $l_{\mathrm{opt}} = 10^3\,a_0$ for the least bound state $v_1=-1,\; \Omega = 0$, under the assumption that the molecules in this state only decay radiatively with the rate $\gamma_1 = 2 \gamma_{\mathrm{atom}}$. These results lead to a promising prospect for the implementation of low-loss optical Feshbach resonances \cite{ciu05} for atom species with a very narrow intercombination line like calcium, i.e. an optically induced change of the scattering length $a_{\mathrm{opt}} \approx \pm~ 300~a_0$ can be achieved at a detuning of about 30 times the natural linewidth, effectively reducing the scattering length close to zero or even negative values.

------------------  

\begin{acknowledgments}
We thank Max Kahmann for experimental support in the early stage of this work, Andreas Koczwara for technical support with the phase-lock of the second spectroscopy laser setup and E. Tiesinga for directing us to the reference regarding the calculation of the retardation in calcium pair interaction.
This work was supported by Deutsche Forschungsgemeinschaft (DFG) through the {\it Center of Quantum Engineering and Space-Time Research (QUEST)} of the Leibniz Universit\"at Hannover, through the Research Training Group 1729 {\it Fundamentals and Applications of Ultra Cold Matter} and CRC 1227 {\it Designed Quantum States of Matter}. 
\end{acknowledgments}

\bibstyle{prstyle}

\end{document}